\documentclass[sigconf]{acmart}

\usepackage{graphicx}
\usepackage{subcaption}
\usepackage{float}
\usepackage{microtype}
\usepackage{stfloats} % For figure* with float option in two-column layout
\usepackage{booktabs}
\usepackage{booktabs}

\AtBeginDocument{%
  }

\acmConference[EWSN'24]{Make sure to enter the correct
  conference title from your rights confirmation emai}{December 10--13 2024}{Abu Dhabi, UAE}

\renewcommand\footnotetextcopyrightpermission[1]{} % Removes footnote with permission statement

\AtBeginDocument{%
  }

\begin{document}

\title{Poster: Could Large Language Models \\ Perform Network Management?}
%\title{Poster: Network Management using Large Language Models}

\author{Zine el abidine Kherroubi, Monika Prakash, Jean-Pierre Giacalone and Michael Baddeley}
\email{{zine.kherroubi,monika.prakash,jean-pierre.giacalone,michael.baddeley}@tii.ae}

\orcid{1234-5678-9012}
\authornotemark[1]
\affiliation{%
  \institution{Technology Innovation Institute}
  \city{Abu Dhabi}
  \country{UAE}
}

\begin{abstract}
Modern wireless communication systems have become increasingly complex due to the proliferation of wireless devices, increasing performance standards, and growing security threats. Managing these networks is becoming more challenging, requiring the use of advanced network management methods and tools. AI-driven network management systems such as Self-Optimizing Networks (SONs) are gaining attention. On the other hand, Large Language Models (LLMs) have been demonstrating exceptional zero-shot learning and generalization capabilities across several domains. In this paper, we leverage the potential of LLMs with SONs to enhance future network management systems. Specifically, we benchmark the use of various LLMs such as GPT-4, Llama, and Falcon, in a zero-shot setting based on their real-time network configuration recommendations. Our results indicate promising prospects for integrating LLMs into future network management systems.

%Modern telecommunication networks require efficient management systems to ensure high reliability and quality of service. However, managing these networks is becoming increasingly challenging due to the growing scalability and complexity of telecommunication protocols and infrastructure. To overcome this challenge, Large Language Models (LLMs) offer promising potential for revolutionizing future network management systems. Indeed, LLMs have demonstrated exceptional zero-shot learning and generalization capabilities across several domains. In this paper, we benchmark the use of LLMs for network configuration recommendations. For that, we evaluate the performance of different widely used LLMs such as GPT-4, Llama, and Falcon in a zero-shot learning setting. Our results indicate promising prospects for integrating LLMs into future network management systems.

%- Set the network management importance.\\
%- Set the network management challenges.\\
%- Set the capabilities of LLM in general.\\
%- Say that we are aiming to see how to use LLM for network management.\\
%- Define the contribution of the paper: benchmarking and perspective (need for fine-tuning).

\end{abstract}



\maketitle

\section{Introduction}
Real-time network management has become a necessity due to the increased complexity, scale and diversity of modern wireless communication systems. The integration of 5G, IoT, AI, edge computing, and SDN/NFV has made real-time decision-making and dynamic adjustments critical to maintaining the performance, security, and efficiency of wireless networks. Most of the existing network management methods are often reactive addressing problems after their occurrence. However, with recent advancements in AI and Machine Learning (ML), proactive and predictive network management techniques such as Self Optimizing Networks (SONs) are gaining attention \cite{AI_net,AI_driven_net}.

%Telecommunication network management is critical for ensuring the reliability, efficiency, and security of communication systems. It facilitates real-time monitoring, fault detection, and resource optimization, thereby minimizing system downtime and enhancing service quality~\cite{stallings2019foundations}. Efficient network management is also essential for handling the increasing data traffic and mitigating cyber threats, as modern networks require higher levels of performance and resilience~\cite{forouzan2021data}. Current network management techniques rely on manual configurations and rule-based systems, which can be both time-consuming and error-prone. While these methods offer a certain level of control, they struggle to keep pace with the growing scale and dynamic nature of modern networks. These conventional approaches often lack the adaptability required to manage highly heterogeneous environments or to respond to emerging cyber threats in real-time. \textcolor{red}{@Monika: you can check and improve this part on network management technics+Limitations.} Recent advancements in AI and Machine Learning (ML) have attracted considerable attention for network management~\cite{AI_enabled_network_1,AI_enabled_network_2}. However, their use in network management is still challenging due to their limited generalization, explainability issues, and high sensitivity to noise. 

Recently, Large Language Models (LLMs) have demonstrated exceptional reasoning and generalization capabilities, which open doors to widespread applications beyond natural language processing tasks. LLMs are now being used in diverse fields such as energy~\cite{llm_energy}, finance~\cite{LLM_finance}, and transportation~\cite{llm_transport}. In ~\cite{telecomGPT}, a telecom-specific LLM, TelecomGPT, was fine-tuned to perform different tasks, such as mathematical modeling and content analysis in the telecoms. An LLM-based Intent translation system~\cite{llm_intent} was also proposed to allows users to express Intents in natural language, and subsequently converts them into Network Service Descriptors (NDSs). To demonstrate how LLMs can simplify and automate complex network management tasks, authors of~\cite{llm_NetConfEval} developed a model-agnostic network configuration benchmark for LLMs called NetConfEval. While these works emphasis the promising perspectives of using LLMs in wireless communication, they remain largely focused on task of natural language understanding and processing. We propose that the real opportunity lies in expanding the role of LLMs as autonomous decision-makers and planners for network management, thus offering a more adaptive, proactive approach to managing modern wireless networks. We therefore propose a benchmark study where we will use and compare various pre-trained LLMs, and instruct them as a zero-shot learners to provide network configuration recommendations.
%Importance of network management (1 to 2 sentences).\\
%Current network management technics.\\
%Limitation of current network management technics.\\
%Importance of AI in general and gen-AI specifically.\\
%Related LLM-based works.\\
%Contribution of the paper.\\

\section{System model}
\begin{figure}[h]
\includegraphics[width=8cm]{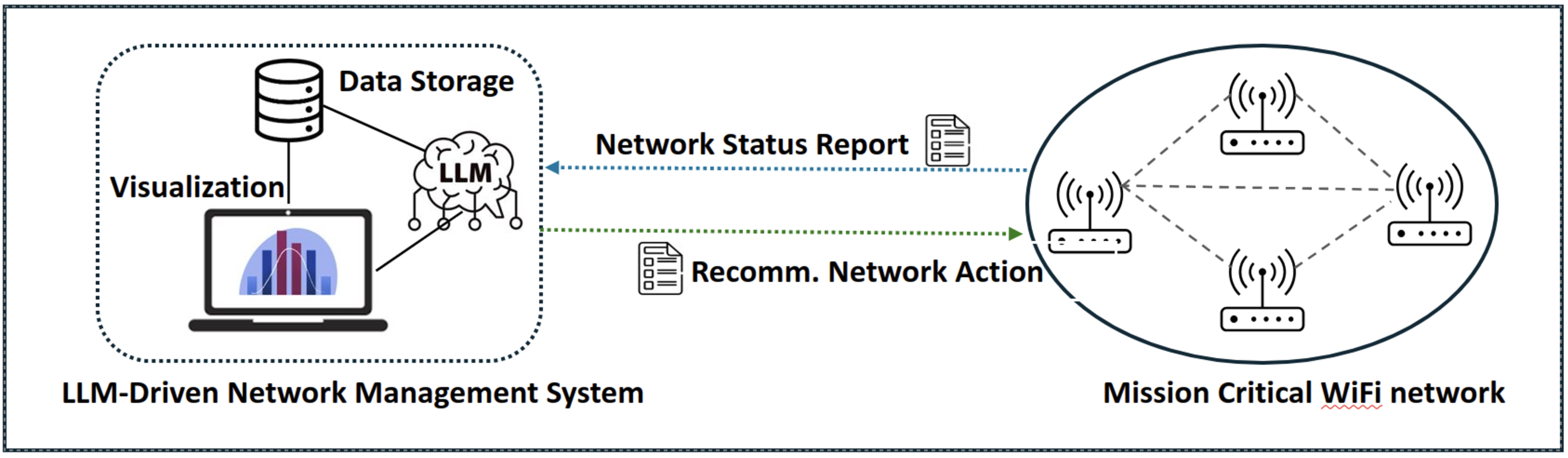}
\caption{System Model}
\vspace{-0.4cm}
\label{system_model}
\end{figure}

We address a scenario of a mission critical network that supports essential services like smart healthcare, smart manufacturing and first responders -- where real-time network management is crucial due to the high stakes involved in maintaining continuous, secure, and reliable communication. Specifically, Wi-Fi based infrastructure-less mesh networks are considered.
%as they are highly valued in mission critical environments for their scalability and coverage. 
As shown in Fig.~\ref{system_model}, each node in the mesh network shares periodic network status reports, which includes metrics such as TX/RX throughput, latency, packet loss, and neighbor nodes. Additionally, they report events like channel interference and jamming detection, which require immediate action to be taken. Based on the observed network states, the LLM is instructed to perform zero-shot reasoning for network configuration recommendations, as illustrated in Table~\ref{prompt_tab}. To do this, we establish the context for the LLM to act as \textit{an expert in network management}, as defined by \textit{the system prompt}. In addition, \textit{the user prompt} provides network state observations, a list of valid actions, and a step-by-step task description, as also illustrated in Table~\ref{prompt_tab}. The LLM then generates a response by selecting an action, which is marked with the tag \textit{<ACTION>chosen action by the LLM</ACTION>}.
%, as shown in Table~\ref{prompt_tab}.
\vspace{-0.3cm}
%LLM: open-source llm advantages, system prompt, observations, Instruction.\\
\begin{table}[t]
\caption{Prompt template}
\vspace{-0.3cm}
  \label{prompt_tab}
\begin{tabular}{l}
\hline
\textbf{System prompt}\\ \hline
\renewcommand{\arraystretch}{0.6}
\begin{tabular}[c]{@{}l@{}}
{\tiny You are a network monitoring expert, and you monitor a wireless mesh network. When there is a network}\\ 
{\tiny security threat such as malicious traffic, jamming, etc., you need to take a valid action among the valid}\\ 
{\tiny actions set to mitigate it. Sometimes, it will be a network performance related update. For example, when}\\ 
{\tiny best neighbors of a node is received, you need to take action to update the neighbors for efficient routing.}\\
{\tiny The neighbors update format is {[}\textless{}node id\textgreater{}, \textless{}node id\textgreater{}{]}. You also need to keep track of the local position of}\\ 
{\tiny nodes and update them accordingly. The position update is provided as {[}x,y,z{]} coordinates. Regarding the}\\ 
{\tiny network, there are 3 nodes on the mesh network named node1, node2 and node3. The mesh network is set}\\
{\tiny to communicate on channel 36 to start. Based on the network observations that you will receive, you are}\\
{\tiny required to choose the best action from the valid action set to keep up the  performance of network and to}\\
{\tiny protect it against security threats. Please, answer that you understood the context.}
\end{tabular}\\ \hline
\textbf{User prompt (Instructions)}
\\ \hline
\renewcommand{\arraystretch}{0.6}
\begin{tabular}[c]{@{}l@{}}{\tiny The network observations are: Network Status from Node1 Best Neighbors List is [2, 3].}\\ {\tiny The valid actions set contains (\#):} \\ {\tiny\# Disconnect all nodes from node 2 \# Disconnect all nodes from node 3 \# Switch all nodes to channel 36} \\ {\tiny\# Switch all nodes to channel 37 \# Switch all nodes to channel 38 \# Switch all nodes to channel 39} \\ {\tiny\# Switch all nodes to channel 40 \# Switch all nodes to channel 41 \# Switch all nodes to channel 42} \\ {\tiny\# Switch all nodes to channel 43 \# Switch all nodes to channel 44 \# Switch all nodes to channel 45} \\ {\tiny\# Switch all nodes to channel 46 \# Update Neighbors of node 1 \# Update Neighbors of node 2} 
\\ {\tiny\# Update Neighbors of node 3 \# Set Network Throughput to 0.1 Mb/s for all nodes} \\ {\tiny\# Set Network Throughput to 2 Mb/s for all nodes \# Set Network Throughput to 10 Mb/s for all nodes}\\ {\tiny\# Update Local Position of node 1 \# Update Local Position of node 2 \# Update Local Position of node 3}\\{\tiny\# No Action.}\\{\tiny INSTRUCTIONS:}\\ {\tiny- You MUST choose only ONE action from the valid action set.}\\ {\tiny- You MUST identify your chosen action by the tag \textless{}ACTION\textgreater{}your choosen action\textless{}/ACTION\textgreater{}.} \\ {\tiny- Do NOT respond with any other additional text, and you CANNOT decline to take an action.}
\end{tabular} 
\\ \hline
\textbf{LLM response}                                     \\ \hline
{\tiny\textless{}ACTION\textgreater{}Update Neighbors of node 1\textless{}/ACTION\textgreater{}}                    \\ \hline
\end{tabular}
\vspace{-0.4cm}
\end{table}
\vspace{-0.2cm}
\section{Benchmarking Results} 
\begin{figure}[t]
\centering
\subfloat[ROUGE-1 score]{\includegraphics[scale=0.27]{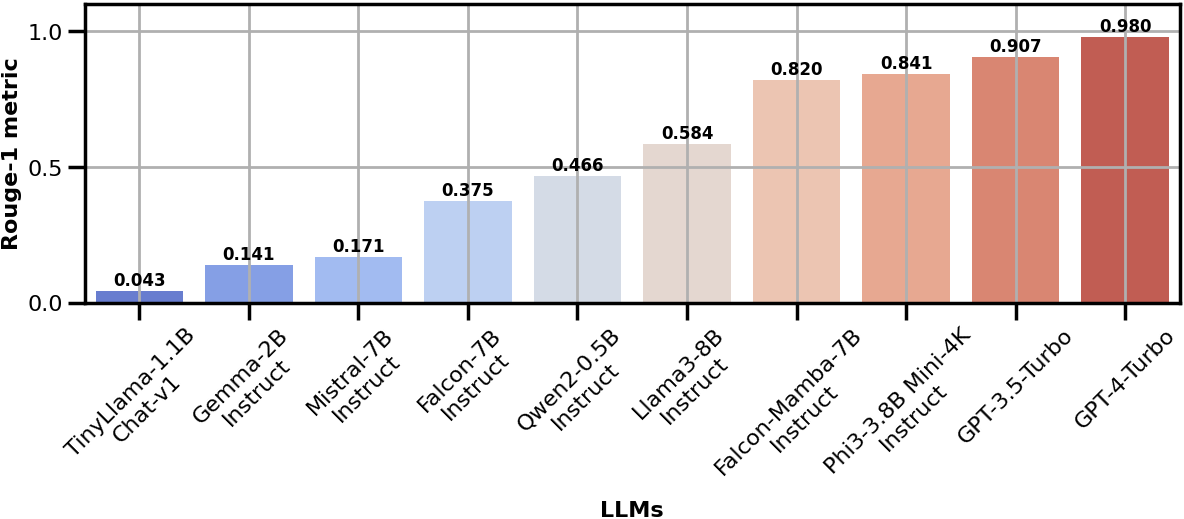}%
\label{bench_rouge_1}}
\vspace{0.05cm} % Adjust the vertical space if necessary
\subfloat[METEOR score]{\includegraphics[scale=0.27]{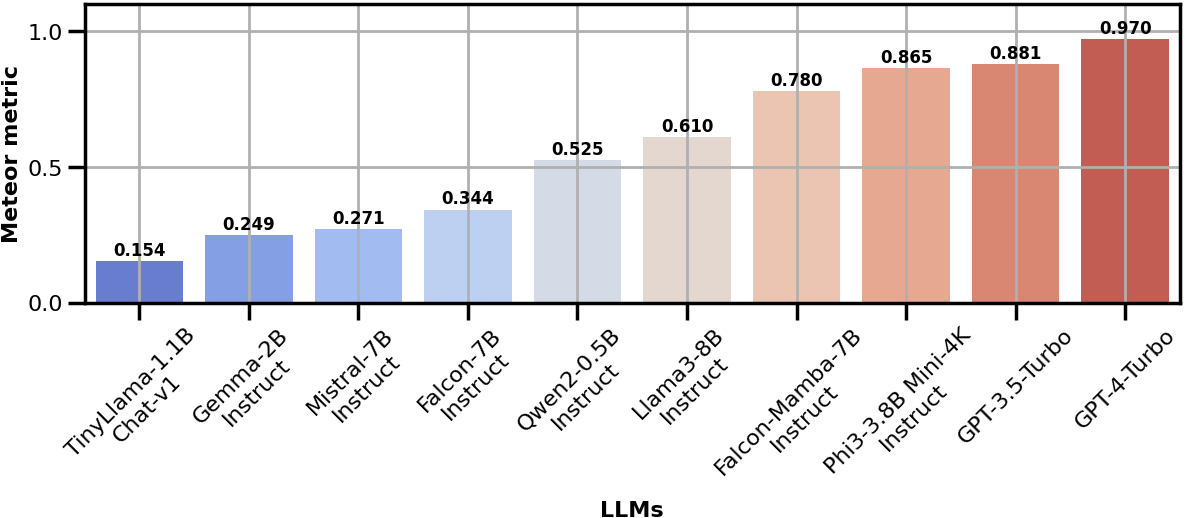}%
\label{bench_meteor}}
\vspace{0.05cm} % Adjust the vertical space if necessary
\subfloat[BLEU score]{\includegraphics[scale=0.27]{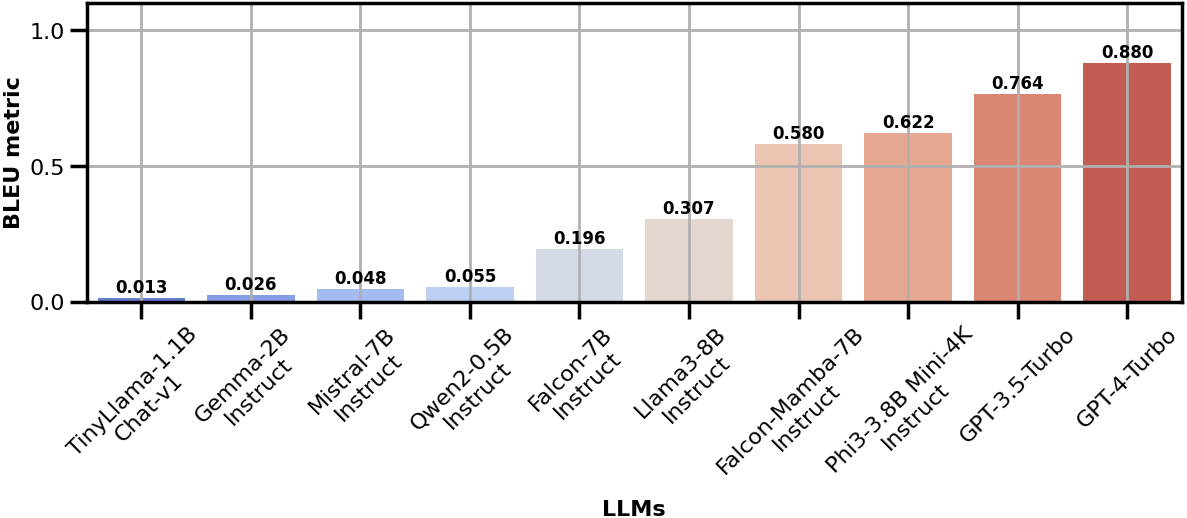}%
\label{bench_bleu}}
\vspace{-0.4cm}
\caption{Benchmarking results across standard LLM metrics.}
\vspace{-0.5cm}
\label{benchmark_results}
\end{figure}

To assess the performance of LLMs for network management, we prompted various models and compared the quality of their generated responses. Our analysis included state-of-the-art open-source models such as the Llama and Falcon series, as well as GPT models accessed via OpenAI APIs. The evaluation was based on three metrics: \textit{ROUGE-1}, \textit{METEOR}, and \textit{BLEU} scores. These metrics were calculated by comparing the LLM-generated responses to preferred labeled responses for each network state. The results are presented in Fig.~\ref{benchmark_results}. Our findings show that GPT models (specifically \texttt{GPT-3.5} and \texttt{GPT-4} \texttt{Turbo}) consistently produce high-quality responses to the instructed tasks, even with zero-shot prompting. This performance can be attributed to their large-scale pre-training and extensive knowledge base. On the open-source side, most models demonstrated lower performance on the task despite well-formatted \textit{system} and \textit{user} prompts. This is likely due to their smaller scale per-training and size.
Interestingly, some recent open-source models, such as \texttt{Phi3-3.8B Mini-4K} and \texttt{Falcon-Mamba-7B}, achieved notably higher performance comparable to the GPT models. Furthermore, the contrast between their high \textit{ROUGE-1} and \textit{METEOR} scores and their average \textit{BLEU} score suggests that while there is some discrepancy with the exact match to the preferred labeled responses, these models grasp the core task and context. Indeed, as demonstrated in Table~\ref{prompt_results} for \texttt{Falcon-Mamba-7B}, the quality of responses from these open-source LLMs is sensitive to minor changes in prompt format. Therefore, further fine-tuning and alignment are necessary to improve their performance on the mesh network management task.
\vspace{-0.20cm}

%performance metrics. (Rouge, Meteor, BLUE, CPU, memory, inference time)\\
%Results (3 figures)\\
%Results discussions:\\
%- We choosed LLM from different suppliers (llama, microsoft, falcon, gpt, etc). This is not an exhaustive study. we selected only LLM < 8B to facilitate embedding them, except GPT who can be accessed using API.\\
%- Very small llm gives bad results.\\
%- Phi3 and Falcon-mamba gives acceptable results.\\
%- GPT performs the best but it need API for access and data privacy is not guarenteed.\\
%- Conclude that we need fine-tune.
%- conclude that we used text-generation metrcis but we can use more quantitative mesure (latency, throughput)

%\begin{table}[h]
%\caption{Sensitivity of Falcon-Mamba-7B to prompt format}
%  \label{prompt_results}
%\begin{tabular}{|l|l|l|l|}
%\hline
%\textbf{Falcon-Mamba-7B}                                       & \textbf{ROUGE-1} & \textbf{METEOR} & \textbf{BLEU} \\ \hline
%\textbf{Prompt ends with \textit{‘\textbackslash{}n’}}                  & 0.82             & 0.78            & 0.58          \\ \hline
%\textbf{Prompt ends without \textit{‘\textbackslash{}n’}}               & 0.38             & 0.38            & 0.54          \\ \hline
%\textbf{Prompt ends with \textit{‘\textbackslash{}n\textbackslash{}n’}} & 0.67             & 0.66            & 0.51          \\ \hline
%\end{tabular}
%\end{table}

\begin{table}[h]
\caption{Sensitivity of Falcon-Mamba-7B to prompt format}
\vspace{-0.3cm}
  \label{prompt_results}
\begin{tabular}{l c c c}
\toprule
\textbf{Falcon-Mamba-7B}                                       & \textbf{ROUGE-1} & \textbf{METEOR} & \textbf{BLEU} \\ 
\midrule
\textbf{Prompt ends with \textit{‘\textbackslash{}n’}}                  & 0.82             & 0.78            & 0.58          \\ 
\textbf{Prompt ends without \textit{‘\textbackslash{}n’}}               & 0.38             & 0.38            & 0.54          \\ 
\textbf{Prompt ends with \textit{‘\textbackslash{}n\textbackslash{}n’}} & 0.67             & 0.66            & 0.51          \\ 
\bottomrule
\end{tabular}
\end{table}
\vspace{-0.4cm}

\section{Conclusion}
In this paper we have explored the use of LLMs for real-time network management systems. Despite their high performance, the use of GPT models for network management tasks presents significant challenges due to limited access, high usage costs, and privacy concerns. However, our benchmarking results clearly indicate that some recent open-source models, such as \texttt{Phi3-3.8B Mini-4K} and \texttt{Falcon-Mamba-7B}, offer promising perspectives, thanks to their smaller size, easy accessibility, and impressive zero-shot performance. Nevertheless, to meet the strict requirements for reliability and resiliency in wireless network management, these open-source models will require further fine-tuning and alignment to be fully effective for this task.
\vspace{-0.1cm}

\bibliographystyle{unsrt}
\bibliography{bibliography}

\end{document}